\documentclass[authoryear,12pt]{elsarticle}

\usepackage{amssymb}
\usepackage{subfigure}

\usepackage[usenames]{color}
\usepackage{algorithmic}
\usepackage{algorithm}
\usepackage{rotating}
\usepackage{amsmath}
\usepackage[dvipdfm,colorlinks,urlcolor=black,linkcolor=black,anchorcolor=black,citecolor=black]{hyperref}
\usepackage[labelfont=bf]{caption} 
\usepackage{longtable,lscape}
\usepackage{amsthm}

\journal{Elsevier}

 \normalsize

\begin{document}

\begin{frontmatter}



\title{A Tabu Search/Path Relinking Algorithm to Solve the Job Shop Scheduling Problem}

\author[HUST]{Bo Peng}

\author[HUST,PolyU]{Zhipeng L\"u\corref{cor}}
\ead{zhipeng.lv@hust.edu.cn}

\author[PolyU]{T.C.E. Cheng}
\ead {Edwin.Cheng@polyu.edu.hk}

\cortext[cor]{Corresponding author.}

\address[HUST]{SMART, School of Computer Science and Technology, Huazhong University of Science and Technology, Wuhan, 430074, P.R. China}
\address[PolyU]{Department of Logistics and Maritime Studies, The Hong Kong Polytechnic University, Hung Hom, Kowloon, Hong Kong}

\begin{abstract}
We present an algorithm that incorporates a tabu search procedure into the framework of path relinking to tackle the job shop scheduling problem (JSP). This tabu search/path relinking (TS/PR) algorithm comprises several distinguishing features, such as a specific relinking procedure and a reference solution determination method. To test the performance of TS/PR, we apply it to tackle almost all of the benchmark JSP instances available in the literature. The test results show that TS/PR obtains competitive results compared with state-of-the-art algorithms for JSP in the literature, demonstrating its efficacy in terms of both solution quality and computational efficiency. In particular, TS/PR is able to improve the upper bounds for 49 out of the 205 tested instances and it solves a challenging instance that has remained unsolved for over 20 years.
\end{abstract}

\end{frontmatter}

\section{Introduction}
\label{Sec_Intro}
The job shop scheduling problem (JSP) is not only one of the most notorious and intractable NP-hard problems, but also one of the most important scheduling problems that arise in situations where a set of activities that follow irregular flow patterns have to be performed by a set of scarce resources. In job shop scheduling, we have a set $M = \{1,\ldots,m\}$ of $m$ machines and a set $J = \{1,\ldots,n\}$ of $n$ jobs. JSP seeks to find a feasible schedule for the operations on the machines that minimizes the makespan (the maximum job completion time), i.e., $C_{max}$, which means the completion time of the last completed operation in the schedule. Each job $j \in J$ consists of $n_{j}$ ordered operations $O_{j,1},\ldots,O_{j,n_{j}}$, each of which must be processed on one of the $m$ machines. Let $O = \{0,1,\ldots,o,o+1\}$ denote the set of all the operations to be scheduled, where operations $0$ and $o + 1$ are dummies, have no duration, and represent the initial and final operations, respectively. Each operation $k\in O$ is associated with a fixed processing duration $P_{k}$. Each machine can process at most one operation at a time and once an operation begins processing on a given machine, it must complete processing on that machine without preemption. In addition, let $p_{k}$ be the predecessor operation of operation $k \in O$. Note that the first operation has no predecessor. The operations are interrelated by two kinds of constraints. First, operation $k \in O$ can only be scheduled if the machine on which it is processed is idle. Second, precedence constraints require that before each operation $k \in O$ is processed, its predecessor operation $p_{k}$ must have been completed.

Furthermore, let $S_{o}$ be the start time of operation $o$ ($S_{0} = 0$). JSP is to find a starting time for each operation $o\ \in O $. Denoting $E_{h}$ as the set of operations being processed on machine $h \in M$, we can formulate JSP as follows:

\begin{equation}
\label{Define_Cmax}
\centering{Minimize\ \ \ C_{max}= \max\limits_{k\in O}}\{S_{k}+P_{k}\},
\end{equation}
subject to
\begin{equation}
\label{Define_Si}
S_{k} \geq 0;\ \ \ \ k= 0,\ldots,o+1,
\end{equation}
\begin{equation}
\label{Define_Si2}
S_{k} - S_{p_{k}} \geq P_{p_{k}};\ \ \ \ k= 1,\ldots,o+1,
\end{equation}
\begin{equation}
\label{Define_Si3}
S_{i}-S_{j} \geq P_{i}\ \ \ or\ \ \ S_{j}- S_{i} \geq P_{j};\ \ \ (i,j) \in E_{h},\ h\in M.
\end{equation}

In the above problem, the objective function (\ref{Define_Cmax}) is to minimize the makespan. Constraints (\ref{Define_Si}) require that the completion times of all the operations are non-negative. Constraints (\ref{Define_Si2}) stipulate the precedence relations among the operations of the same job. Constraints (\ref{Define_Si3}) guarantee that each machine can process no more than one single operation at a time.

Over the past few decades, JSP has attracted much attention from a significant number of researchers, who have proposed a large number of heuristic and metaheuristic algorithms to find optimal or near-optimal solutions for the problem. One of the most famous algorithms is the tabu search (TS) algorithm TSAB proposed by \cite{Nowicki1996TSAB}. \cite{Nowicki2005i-TSAB} later extend algorithm TSAB to algorithm i-TSAB, which \cite{Beck2011CP/LS} combine with a constraint programming based constructive search procedure to create algorithm CP/LS. 
\cite{Pardalos2006GES} propose algorithm GES, which is based on global equilibrium search techniques. \cite{Zhang2007TS} extend the N6 neighbourhood proposed by \cite{Balas1998GLS} to the N7 neighbourhood and \cite{Zhang2008TSSA} combine TS with SA to create algorithm TS/SA, which outperforms almost all the algorithms. \cite{Nagata2009GES} present a local search framework termed guided ejection search, which always searches for an incomplete solution for JSP. Recently, \cite{Goncalves2013BRKGA} present the biased random-key genetic algorithm BRKGA, which is able to improve the best known results for 57 instances and outperforms all the reference algorithms considered in their paper. From all these algorithms, it is apparent that the recent state-of-the-art algorithms either hybridize several strategies instead of using a single algorithm or employ a population-based algorithm instead of a single-solution based one.

Among the metaheuristic approaches used to tackle JSP, especially most of the state-of-the-art algorithms for JSP, a powerful local search procedure is always necessary. As one of the most popular local search algorithms, TS has been widely used by researchers to tackle JSP, e.g., \cite{Nowicki2005i-TSAB}, \cite{Zhang2007TS}, \cite{Nasiri2012GES/TS}, \cite{Shen201214}, and \cite{Goncalves2013BRKGA}, among others.

On the other hand, \cite{Aiex2003GRASPwithPR} apply path relinking within a GRASP procedure as an intensification strategy to tackle JSP. Furthermore, \cite{Nowicki2005i-TSAB} improve their famous algorithm TSAB by introducing a new initial solution generator based on path relinking. Recently, \cite{Nasiri2012guidedTSPR} apply i-TSAB to tackle JSP using the N1 neighbourhood as the path relinking procedure.


The above observations and considerations motivate us to develop a more robust algorithm for JSP via combining the more global relinking approach and the more intensive TS, which consists of several distinguishing features. In this vein, we design the tabu search and path relinking (TS/PR) algorithm, which is able to strike a better balance between the exploration and exploitation of the search space in a flexible manner. We summarize the main contributions of TS/PR as follows:

\begin{itemize}
\item Compared with the state-of-the-art algorithms for tackling JSP, TS/PR consists of several distinguishing features. In particular, it uses a specific mechanism to adaptively construct the path linking the initiating solution and the guiding solution, as well as using two kinds of improvement method to determine the reference solution.

\item We test the performance of TS/PR by applying it to solve 205 benchmark JSP instances widely used in the literature. The test results show the efficacy TS/PR in terms of both solution quality and computational efficiency. In particular, TS/PR is able to improve the upper bounds for 49 out of the 205 tested instances in a reasonable time and it finds the optimal solution for the challenging instance SWV15, which has remained unsolved for over 20 years.
\end{itemize}

The remaining part of the paper is organized as follows: Section \ref{Sec TS/PR} describes in detail the components of TS/PR. Section \ref{sec results Experiment3} presents the detailed computational results and comparisons between TS/PR and some best performing algorithms in the literature for tackling six sets of a total of 205 challenging benchmark JSP instances. Finally, we conclude the paper and suggest future research topics in Section \ref{sec conclusions}.

\section{The TS/PR Algorithm}
\label{Sec TS/PR}

\subsection{Main Framework}
\label{subsec sol main scheme}

In principle, TS/PR repeatedly operates between a path relinking method that is used to generate promising solutions on the trajectory set up from an initiating solution to a guiding solution, and a TS procedure that improves the generated promising solution to a local optimum. Algorithm \ref{algorithm_outline} presents the main procedure of TS/PR.

\begin{algorithm}[!hp]
\begin{scriptsize}
 \caption{Outline of algorithm TS/PR for JSP}\label{algorithm_outline}
 \begin{algorithmic}[1]
   \STATE \sf \textbf{Input}: $J$, $M$, and $P_{k}$
   \STATE \textbf{Output}: $C_{max}$ and the best solution $S^*$ found so far
   \STATE $P=\{S^{1},\ldots,S^{p}\}$ $\leftarrow$
   Population\_Initialization() \hfill /$*$ Section \ref{subsec initial} $*$/
   \FOR{$i=\{1,\ldots,p\}$}
       \STATE $S^{i}$ $\leftarrow$ Tabu\_Search($S^{i}$) \hfill /$*$ Section \ref{subsec TS} $*$/
   \ENDFOR
   \STATE $S^{*}=arg \ min \{f(S^{i})\ |\ i=1,\ldots,p\}$
   \STATE \textsl{PairSet} $\leftarrow$  $\ \{(S^{i},S^{j})\  |\ S^{i} \in P,\ S^{j} \in P$ and $S^{i}\neq S^{j}\}$
   \REPEAT
        \STATE Randomly select one solution pair \{$S^{i}$,$S^{j}$\} from \textsl{PairSet}
        \STATE $S^{p+1}$ $\leftarrow$ Path\_Relinking($S^{i},S^{j}$)\ ,\ $S^{p+2}$ $\leftarrow$ Path\_Relinking($S^{j},S^{i}$)\hfill /$*$ Section\ref{subsec PR} $*$/
        \STATE $S^{p+1}$ $\leftarrow$ Tabu\_Search($S^{p+1}$), $S^{p+2}$ $\leftarrow$ Tabu\_Search($S^{p+2}$)\hfill /$*$ Section \ref{subsec TS}
        $*$/
        \IF{$S^{p+1}$(or $S^{p+2}$) is better than $S^*$}
           \STATE $S^*= S^{p+1}$(or $S^{p+2}$)
        \ENDIF
        \STATE Tentatively add $S^{p+1}$ and $S^{p+2}$ to population $P$:
        $P'=P\cup \{S^{p+1},S^{p+2}\}$
        \STATE \textsl{PairSet} $\leftarrow$ \textsl{PairSet} $\cup \ \{(S^{p+1},S^{k})\  |\ S^{k} \in P$ and $ S^{k} \neq S^{p+1} \}$
        \STATE \textsl{PairSet} $\leftarrow$ \textsl{PairSet} $\cup \ \{(S^{p+2},S^{k})\  |\ S^{k} \in P$ and $S^{k} \neq S^{p+2} \}$
        \STATE Identify the two worst solutions $S^{u}$ and $S^{v}$ in the temporary population $P'$\\
        \STATE Generate new population by removing the two worst solutions $S^{u}$ and $S^{v}$:\\
        $P=\{S^{1},\ldots,S^{p},S^{p+1},S^{p+2}\}\backslash \{S^{u},S^{v}\}$
        \STATE Update \textsl{PairSet}:\\
          \ \ \ \ \textsl{PairSet} $\leftarrow$ \textsl{PairSet} $\backslash \ \{(S^{u},S^{k})\  |\ S^{k} \in P$ and $S^{k} \neq S^{u} \}$\\
          \ \ \ \ \textsl{PairSet} $\leftarrow$ \textsl{PairSet} $\backslash \ \{(S^{v},S^{k})\  |\ S^{k} \in P$ and $S^{k} \neq S^{v} \}$
   \UNTIL{a stop criterion is met}
 \end{algorithmic}
 \end{scriptsize}
\end{algorithm}

\subsection{Initial Population}
\label{subsec initial}

 In TS/PR, the initial population is constructed as follows: Starting from scratch, we randomly generate a feasible solution and then optimize the solution to become a local optimum using our improvement method (see Section \ref{subsec TS}). The resulting improved solution is added to the population if it does not duplicate any solution currently in  the population. This procedure is repeated until the size of  the population
reaches the cardinality $p$.

\subsection{Tabu Search Procedure}
\label{subsec TS}


Our TS procedure uses the N7 neighbourhood proposed by \cite{Zhang2007TS}. It stops if the optimal solution is found or the best objective value has not been improved for a given number of TS iterations, called the tabu search \textsl{cutoff}. The interested reader may refer to the hybrid evolutionary algorithm HEA presented in \cite{Cheng2013HEA} for more details.

\subsection{Path Relinking Procedure}
\label{subsec PR}

The relinking procedure is used to generate new solutions by exploring trajectories (confined to the neighbourhood space) that connect high-quality solutions. The solution that begins the path is called the initiating solution while the solution that the path leads to is called the guiding solution. The $PathSet$ is a list of candidate solutions that stores all the solutions generated during the path relinking procedure. After the relinking procedure, a so-called reference solution is chosen from the $PathSet$ that serves to update the population. In order to better describe the relinking procedure, we give some definitions in Table \ref{symbol description}.

\begin{table}[!h]
\begin{scriptsize}
\caption{The description of the symbols used in TS/PR}
\label{symbol description}
\scalebox{0.8}{
\begin{tabular}{p{2cm}p{14cm}}
\hline
Symbols           & Description  \\
\hline

$j_{k,i}$         & The $i$th operation executed on machine $k$. \\

$S$               & A schedule for JSP is represented by permutations of operations on the machines:
$ \{(j_{1,1},j_{1,2},\ldots,j_{1,n}),(j_{2,1},j_{2,2},\ldots,j_{2,n}),\ldots,(j_{m,1},j_{m,2},\ldots,j_{m,n})\}$. \\

$S^I$             & The initiating solution for the relinking procedure.   \\

$S^G$             & The guiding solution in the relinking procedure.  \\

$S^C$             & The current solution during the relinking procedure.\\

$CS(S^{I},S^{G})$ & The common sequence between $S^{I}$ and $S^{G}$: $ \{j^{I}_{k,i}|j^{I}_{k,i} = j^{G}_{k,i}, k\in M,i\in N\}$. \\

$NCS(S^{I},S^{G})$& The set of elements not in the common sequence between $S^{I}$ and $S^{G}$:
$ \{j^{I}_{k,i}|j^{I}_{k,i}\neq j^{G}_{k,i}, k\in M,i\in N\}$. \\

$Dis(S^{I},S^{G})$& The distance between $S^{I}$ and $S^{G}$:
$ |NCS(S^{I},S^{G})|$.    \\

\textsl{PairSet}  & A set that stores the candidate solution pairs for path building. \\

\textsl{PathSet}  & A set that stores the candidate solutions on a single path that will be optimized by the improvement method.  \\

$\alpha$          & The minimum distance between the initiating (or guiding) solution and the first (or last) solution in the \textsl{PathSet}. \\

$\beta$           & The interval for choosing the candidate solutions in \textsl{PathSet}. \\

\hline
\end{tabular}
}
\end{scriptsize}
\end{table}

Contrary to previous studies, our proposed path relinking process mainly integrates two complementary key components to ensure search efficiency. The first one is the constructing approach used for establishing the path between the initiating and the guiding solutions. In the related literature, \cite{Nasiri2012guidedTSPR}'s relinking swaps adjacent operations on a machine, while GRASP/PR by \cite{Aiex2003GRASPwithPR} swaps different operations on each machine in turn. However, in this study we swap two different operations on one machine randomly, where both the operations and the responding machines are randomly chosen. More details will be presented in Section \ref{Sec TS/PR}. The second one is the method used to choose the reference solution. In related studies, \cite{Aiex2003GRASPwithPR} simply consider the solution with the best makespan in the path as the reference one, while \cite{Nasiri2012guidedTSPR} follow \cite{Nowicki2005i-TSAB}'s i-TSAB whereby it goes from the initiating solution, then stops at a specific iteration and returns the current solution as the reference solution. In contrast, we devise a dedicated strategy based on the adaptive distance-control mechanism to obtain the most promising solution. Therefore, the path relinking approach plays the important role of diversification in coordinating with the efficient TS procedure. 

\subsubsection{The relinking procedure:}
\label{subsubsec relinking}
\begin{algorithm}[!h]
\begin{scriptsize}
 \caption{Pseudo-code of the relinking procedure}\label{algorithm_PR}
 \begin{algorithmic}[1]
   \STATE \sf \textbf{Input}: Initiating solution $S^{I}$ and guiding solution $S^{G}$
   \STATE \textbf{Output}: A reference solution $S^{R}$
   \STATE  Identify the elements not in common sequence between $S^{I}$ and $S^{G}$, denoted as $NCS(S^{I},S^{G})$
   \STATE $S^{C}=S^I$, $PathSet={\O}$
   \STATE //Lines 6-10: Change $S^{C}$ to $S^G$ by $\alpha$ consecutive swap moves /$\ast$Section \ref{subsubsec_path_construction}$\ast$/
   \FOR{$k=\{1,\ldots,\alpha\}$}
       \STATE  Randomly select an element $S^{G}_{i}$  in $NCS(S^{C},S^{G})$
       \STATE  Swap element $S^{G}_i$ and another one in $S^{C}$ such that $S^{G}_i$'s position in $S^{C}$ is the same as that in $S^{G}$
        \STATE $NCS(S^{C},S^{G})$   $\leftarrow$ $ NCS(S^{C},S^{G}) \ \backslash \ S^{G}_i $
   \ENDFOR

   \STATE $PathSet\ \leftarrow\ PathSet\ \cup\ S^C$

   \STATE //Lines 13-20: Construct the path with an interval $\beta$ until its distance to $S^{G}$ is less than $\alpha$  /$\ast$ Section \ref{subsubsec_path_construction}$\ast$/

   \WHILE{$Dis(S^C,S^G) > \alpha$ }
        \FOR{$k=\{1,\ldots,\beta\}$}
       \STATE  Randomly select an element $S_{i}^{G}$ in $NCS(S^{C},S^{G})$
       \STATE  Swap element $S^{G}_i$ and another one in $S^{C}$ such that $S^{G}_i$'s position in $S^{C}$ is the same as in $S^{G}$
       \STATE $NCS(S^{C},S^{G})$   $\leftarrow$ $ NCS(S^{C},S^{G}) \ \backslash \ S^{G}_i $
       \ENDFOR
       \STATE $PathSet\ \leftarrow\ PathSet\ \cup\ S^C$
   \ENDWHILE

   \STATE //Lines 22-30: Choose the reference solution from $PathSet$ \hfill  /$\ast$ Section \ref{subsubsec_path_solution}$\ast$/

   \STATE let $q$ be the cardinality of $PathSet$:\ \ $q = |PathSet|$
   \FOR{$S_k\in PathSet$, $k=\{1,\ldots,q\}$}
       \IF{solution $S_k$ is an infeasible solution}
           \STATE  $S_{k}$ $\leftarrow$  Repair($S_k$)
        \ENDIF

       \STATE $S_{k}$ $\leftarrow$ Tabu\_Search($S_{k}$) with a small number of iteration $si$
   \ENDFOR
   \STATE Record the best solution in $PathSet$ as the reference solution $S^{R}$:\\
          $S^{R}$= arg $min \{f(S_{k}), k=1,\ldots,q\}$
   \STATE $S^{R}$ $\leftarrow$ Tabu\_Search($S^{R}$) with a large number of iteration $li$
   \RETURN $S^{R}$
 \end{algorithmic}
 \end{scriptsize}
\end{algorithm}

Algorithm \ref{algorithm_PR} presents the relinking procedure in detail. Section \ref{subsubsec_path_construction} presents how we construct a path from the initiating solution $S^{I}$ to the guiding solution $S^{G}$. Section \ref{subsubsec_path_solution} explains how we choose a subset of candidate solutions, denoted by the $PathSet$, possibly as the reference solution. In Section \ref{subsubsec_ref_solution} the reference solution is determined by applying both $slight$ and $strong$ TS procedures to the candidate solutions in the $PathSet$.

\subsubsection{Path construction:}
\label{subsubsec_path_construction}

\begin{figure}[!htbp]
  \centering\scalebox{0.66}{\includegraphics{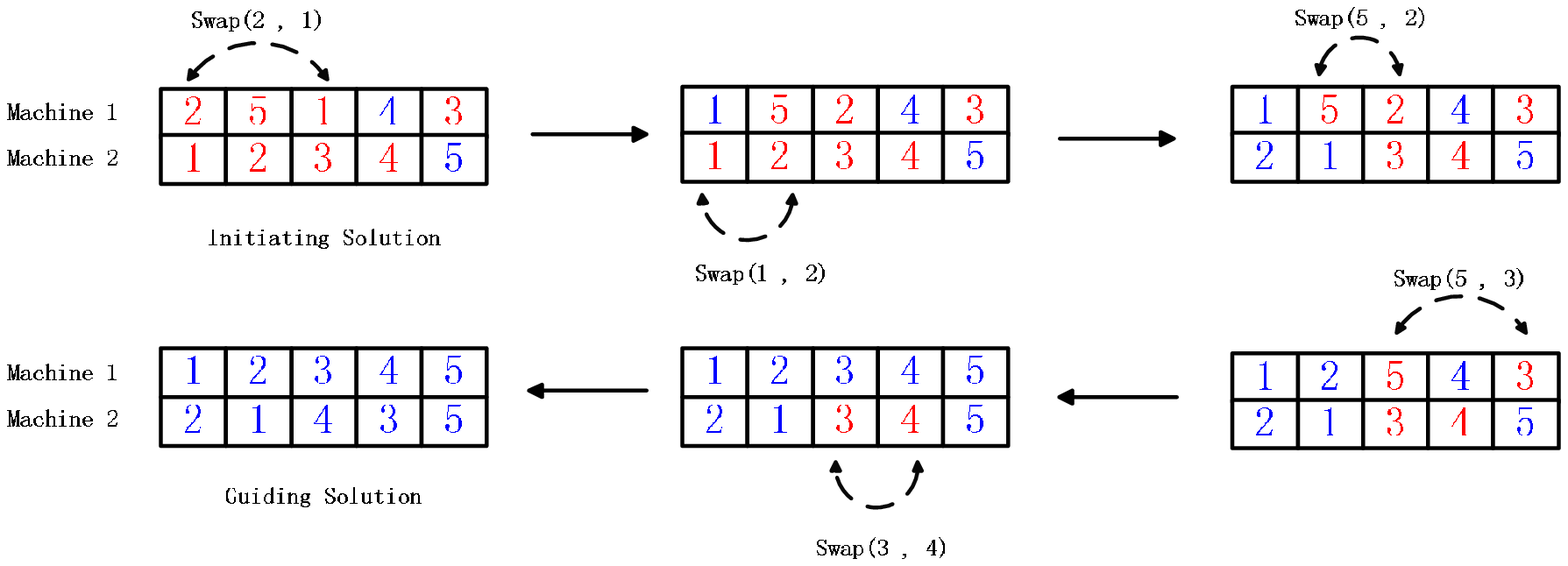}}\\
  \centering\caption{Path construction procedure}\label{fig_pathconstruction}
\end{figure}

We employ the swap operation, which swaps two elements on the same machine, to build a path from the initiating solution $S^{I}$ to the guiding solution $S^{G}$. At the beginning, the current solution $S^{C}$ is assigned as $S^{I}$. In each iteration, $S^{C}$ is changed by a swap operation towards the guiding solution $S^{G}$. Specifically, in $S^{C}$ we iteratively swap two (random) elements that are in different order in $S^{C}$ and $S^{G}$. Figure \ref{fig_pathconstruction} gives an example of executing the path construction procedure on two machines. In this example, the initiating solution $S^{I}$ is transformed into the guiding solution $S^{G}$ in five swap moves. It should be noted that \cite{Aiex2003GRASPwithPR}'s GRASP/PR only swaps the operations on one machine in each iteration. In other words, only if the operations on one machine are the same will the next machine be taken into consideration, whereas in our study both the operations and machines are randomly chosen. Despite this subtle difference, our approach enhances the possibility of constructing a diversified path.

\subsubsection{Path solution selection:}
\label{subsubsec_path_solution}

Since two consecutive solutions on a relinking path differ only by swapping two elements on a machine, it is not productive to apply an improvement method to each solution on the path since many of these solutions would lead to the same local optimum. More importantly, the improvement method is very time consuming, so we should restrict its use to only a subset of promising solutions, denoted as the $PathSet$.

\begin{figure}[!htbp]
  \centering\scalebox{1}{\includegraphics{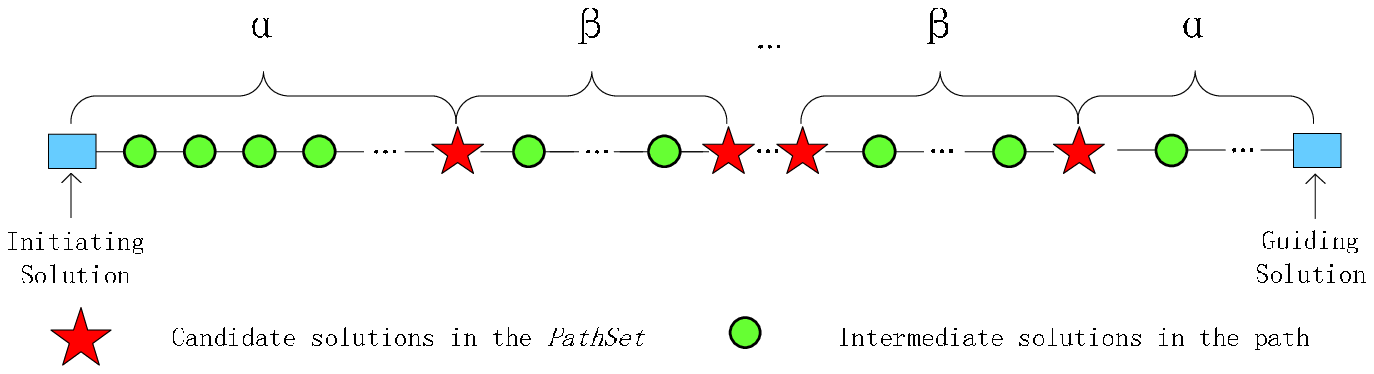}}\\
  \centering\caption{Illustration of the path solution selection procedure}\label{fig_solutionselection}
\end{figure}

We construct the $PathSet$ as follows: First, we choose the first solution at a distance of at least $\alpha$ from the initiating solution in the $PathSet$. Then, for each interval of $\beta$ swap moves, we add a solution to the $PathSet$ until the distance between the current solution and the guiding solution is less than $\alpha$. In such a way, the candidate solution list $PathSet$ is constructed. Figure \ref{fig_solutionselection} gives an illustration of the path solution selection procedure.

\subsubsection{Reference solution determination:}
\label{subsubsec_ref_solution}

As soon as the $PathSet$ is built, we need to determine the reference solution to update the population. For this purpose, we first employ a TS with short iterations (we label it as a $slight$ TS) to optimize each solution in the $PathSet$ to become a local optimum. Then, the best optimized solution is selected and further optimized using a TS with long iterations (we label it as a $strong$ TS). This optimized solution is chosen as the reference solution. The reason is that a slight TS is not too time consuming but can optimize a solution to some extent, from which we can judge which solution is more promising than the others. As long as a reference solution is chosen, it is necessary to optimize it as far as possible, so we utilize a $strong$ TS to optimize it.

It should be noted that it is possible that the solutions in the $PathSet$ are infeasible solutions because of the random swap operations during the path construction procedure, which may violate the precedence constraints. In this case, the previous literature is often inclined to abandon or delete them, e.g., GRASP/PR from \cite{Aiex2003GRASPwithPR}. However, we utilize a special technique proposed by \cite{Qing-dao-er-ji2012HGA} to repair the infeasible solutions to feasible ones (lines 24-26 in Algorithm \ref{algorithm_PR}).

\section{Computational Results}
\label{sec results Experiment3}

In this section we report extensive experimental results of applying TS/PR to tackle six sets of a total of 205 benchmark JSP instances widely used in the literature. We coded TS/PR in C++ and ran it on a PC with a Quad-Core AMD Athlon 3.0GHz CPU and 2Gb RAM under the Windows 7 operating system. Table \ref{Parameter_Settings} gives the descriptions and settings of the parameters used in TS/PR, in which the last column denotes the settings for the set of all the instances. Given the stochastic nature of TS/PR, we solved each problem instance ten times independently. For each run, in view of the different levels of difficulty of the benchmark instances, we set different total time limits for applying TS/PR to tackle them. Table \ref{Time_limit} gives the set time limits.

\begin{table}[!h]
\begin{scriptsize}
\caption{The settings of some important parameters in TS/PR}
\label{Parameter_Settings}
\scalebox{0.8}{
\begin{tabular}{p{2.5cm}p{3cm}p{10cm}}
\hline
Parameter & Value           & Description  \\
\hline

$\alpha$  &$\frac{dis}{5}$   & Minimum distance between solutions in $PathSet$ and $S^{I}$ and $S^{G}$\\

$\beta$   &$max(\frac{dis}{10},2)$& Interval for choosing the path solutions   \\

$si$      & 500                                                         & The number of iterations for the slight tabu search\\

$li$      & 12500                                                       & The number of iterations for the strong tabu search\\

$p$       & 30                                                          & Population size                     \\
\hline
\end{tabular}
}
\end{scriptsize}
\end{table}

\begin{table}[!h]
\begin{scriptsize}
\caption{The settings of the time limit for different categories of instances}
\label{Time_limit}
\scalebox{0.8}{
\begin{tabular}{p{2.5cm}p{2.5cm}p{2.5cm}p{2.5cm}p{2.5cm}p{2.2cm}}
\hline
Instance Name   :        & SWV12(15)     &    DMU56-65 & DMU66-70 &DMU71-80 & Other instances  \\
\hline
Time limit  : & 2 hours & 2 hours & 4 hours   & 5 hours & 1 hour   \\
\hline

\end{tabular}
}
\end{scriptsize}
\end{table}

In this paper we report the detailed results on testing TS/PR (such as the best and relative values, and the time required to obtain the results) to faciliate future comparisons. All these benchmark instances can be downloaded from the OR-Library\footnote{\url{http://people.brunel.ac.uk/~mastjjb/jeb/orlib/jobshopinfo.html}} and Shylo's webpage\footnote{\url{http://plaza.ufl.edu/shylo/jobshopinfo.html}}.

To test the performance of TS/PR, we consider the following well-known JSP instance classes:

\begin{itemize}
  \item The first set of benchmark instances consists of 13 basic instances, including the three instances FT6, FT10, and FT20 due to \cite{Fisher1963FT06-20}, and the ten instances ORB01-10 due to \cite{ApplegateSpring1991ORB01-10};
  \item The second set of benchmark instances consists of the 40 classic instances LA01-40 due to \cite{Lawrence1984LA01-40}.
  \item The third set of benchmark instances consists of the three difficult instances ABZ07-09 due to \cite{Adams1988LA01-40}, and the four instances YN01-04 due to \cite{Yamada1992YN1-4}. Although this instance set is not large, the optimal values of these instances are still unknown.
  \item The fourth set of benchmark instances consists of the 15 instances SWV01-15 due to \cite{Storer1992SWV01-15}.
  \item The fifth set of benchmark instances consists of 50 of the most difficult instances TA01-50 due to \cite{TaillardSpring1994TA01-50}.
  \item The sixth set of benchmark instances consist of 80 of the most difficult instances DMU01-80 due to \cite{Demirkol1997DMU01-80}.
\end{itemize}

To measure the performance of TS/PR, we calculate the relative error (RE) using the relative deviation formula: RE = 100 $\times$ (UB$_{solve}$ - LB$_{best}$)/ LB$_{best}$, for each instance, where LB$_{best}$ is the best known lower bound and UB$_{solve}$ is the best makespan found by all of the tested algorithms. Subsequently, we calculate the mean relative error (MRE) for a given algorithm as the mean RE over all the tested instances.

In our experiments, the best known LB and UB were obtained from the following papers and website pages. Note that these algorithms can generate the upper bounds for almost all of the instances and can be considered as the current state-of-the-art algorithms for JSP. In the context of performance evaluation, we compare TS/PR mainly with these state-of-the-art algorithms in detail, which include: i-TSAB by \cite{Nowicki2005i-TSAB},
GES by \cite{Pardalos2006GES}, TS by \cite{Zhang2007TS}, TS/SA by \cite{Zhang2008TSSA}, AlgFix by \cite{Pardalos2010AlgFix}, CP/LS by \cite{Beck2011CP/LS}, GES/TS by \cite{Nasiri2012GES/TS}, HGA by \cite{Qing-dao-er-ji2012HGA}, BRKGA by \cite{Goncalves2013BRKGA}, Taillard's URL\footnote{\url{http://mistic.heig-vd.ch/taillard/problemes.dir/ordonnancement.dir/jobshop.dir/best_lb_up.txt}}, and Shylo's webpage\footnote{\url{http://plaza.ufl.edu/shylo/jobshopinfo.html}}.

\subsection{Computational results on the first two sets of instances}

\label{subsec Computational Results 1}

\begin{table}[!h]
\begin{scriptsize}
\caption{Computational results and comparisons for the first set of instances (FT series and ORB01-10)}
\label{table FT and ORB}
\scalebox{0.8}{
\begin{tabular}
{p{1.2cm}p{1.3cm}p{1cm} p{1cm}p{1cm}p{1cm} p{0.3cm} p{1cm}  p{1cm} p{1cm}p{0.9cm} p{0.7cm} }
\hline
 & & & \multicolumn{3}{c}{TS/PR}& &\multicolumn{2}{c}{BRKGA} & & &    \\
\cline{4-6} \cline{8-9}
 \centering{\raisebox{2.50ex}[0cm][0cm]{Problem}} & \centering{\raisebox{2.50ex}[0cm][0cm]{Size}} &
 \centering{\raisebox{2.50ex}[0cm][0cm]{$OPT$}}  & Best & $M_{av}$ & $T_{av}$&
 & Best & $M_{av}$ &\raisebox{2.50ex}[0cm][0cm]{TS/SA} &\centering{\raisebox{2.50ex}[0cm][0cm]{GES}}  &\raisebox{2.50ex}[0cm][0cm]{TS}
    \\  \hline
FT06   & $6  \times 6$  &55   &55   &55    &0.03 & &55  &55  &55    &-    &55\\
FT10   & $10 \times 10$ &930  &930  &930   &4.75 & &930 &930 &930   &-    &930\\
FT20   & $20 \times 5$ &1165  &1165 &1165  &0.18 & &1165&1165&1165  &-    &1165\\
ORB01  & $20 \times 20$ &1059 &1059 &1059  &0.51 & &1059&1059&1059  &1059 &-  \\
ORB02  & $20 \times 20$ &888  &888  &888   &1.69 & &888 &888 &888   &888  &-  \\
ORB03  & $20 \times 20$ &1005 &1005 &1005  &1.46 & &1005&1005&1005  &1005 &-  \\
ORB04  & $20 \times 20$ &1005 &1005 &1005  &3.71 & &1005&1005&1005  &1005 &-  \\
ORB05  & $20 \times 20$ &887  &887  &887   &7.28 & &887 &887 &887   &887  &-  \\
ORB06  & $20 \times 20$ &1010 &1010 &1010  &1.81 & &1010&1010&1010  &1010 &-  \\
ORB07  & $20 \times 20$ &397  &397  &397   &0.13 & &397 &397 &397   &397  &-  \\
ORB08  & $20 \times 20$ &899  &899  &899   &3.99 & &899 &899 &899   &899  &-  \\
ORB09  & $20 \times 20$ &934  &934  &934   &0.47 & &934 &934 &934   &934  &-  \\
ORB10  & $20 \times 20$ &944  &944  &944   &0.09 & &944 &944 &944   &944  &-  \\
\hline
MRE    &                &0    &0    & 0    &2.01 & &0   &0   &0     &0    &0  \\
\hline
\end{tabular}
}
\end{scriptsize}
\end{table}

We conducted the first experiment to evaluate the performance of TS/PR in tackling the sets of 53 benchmark JSP instances: FT06, 10, 20, ORB01-10, and LA01-40. The number of operations for these instances ranges from 36 to 300. Tables \ref{table FT and ORB} and \ref{tabel LA} provide a summary of the performance comparisons between TS/PR and BRKGA, TS/SA, GES, and TS for instances of the FT, ORB01-10, and LA classes. In both tables, the column $OPT$ lists the optimal solution for each instance. The following three columns Best, $M_{av}$, and $T_{av}$ show the best makespan, average makespan, and average computing time in seconds to obtain the best value, respectively, by TS/PR over ten runs. The next two column presents the best makespan and average makespan in BRKGA, and the last three columns give the best results of the reference algorithms TS/SA, GES, and TS. The last row presents the MRE value averaged over one set of instances. In addition, the row TS/PR reports the MRE value for part of the instances since some reference algorithms only give results for part of the instances.  For each class of instances, the best MRE (Best), the average MRE ($M_{av}$), and the average running time ($T_{av}$) are listed for each algorithm. Due to their relatively small sizes, most of these instances, except for LA29, are very easy to solve by TS/PR and the reference algorithms.

From Table \ref{table FT and ORB}, we see that TS/PR can easily reach the optima within 2.01 seconds on average for the 13 FT and ORB instances. From Table \ref{tabel LA}, we see that TS/PR can easily reach the optima for all the 40 LA instances, except for LA29, within 13.9 seconds on average. TS/PR is only slightly worse than algorithm GES but is better than the other three algorithms TS/SA, TS, and HGA in terms of solution quality.

\begin{table}[!hp]
\begin{scriptsize}
\caption{Detailed computational results and comparisons for the second set of instances LA01-40}
\label{tabel LA}
\scalebox{0.8}{
\begin{tabular}
{p{1.1cm}p{1.3cm}p{0.9cm} p{0.9cm}p{0.9cm}p{0.9cm} p{0.01cm} p{0.9cm}p{0.9cm}  p{0.9cm}p{0.9cm} p{0.9cm}p{0.9cm} }
\hline
 & & & \multicolumn{3}{c}{TS/PR}& & \multicolumn{2}{c}{BRKGA} & & & &    \\
\cline{4-6} \cline{8-9}
 \raisebox{2.50ex}[0cm][0cm]{Problem} & \raisebox{2.50ex}[0cm][0cm]{Size} &\raisebox{2.50ex}[0cm][0cm]{$OPT$}  & Best & $M_{av}$ & $T_{av}$
 & &  Best & $M_{av}$ &\raisebox{2.50ex}[0cm][0cm]{GES} &\raisebox{2.50ex}[0cm][0cm]{TS/SA} &\raisebox{2.50ex}[0cm][0cm]{TS} &\raisebox{2.50ex}[0cm][0cm]{HGA}
    \\  \hline
 LA01   & $10 \times 5$  & 666    &666&666&0.05   &  & 666   &666    &666  & -    & -  & 666 \\
 LA02   & $10 \times 5$  & 655    &655&655&0.05   &  & 655   &655    &655  & -    & -  & 655 \\
 LA03   & $10 \times 5$  & 597    &597&597&0.06   &  & 597   &597    &597  & -    & -  & 597 \\
 LA04   & $10 \times 5$  & 590    &590&590&0.05   &  & 590   &590    &590  & -    & -  & 590 \\
 LA05   & $10 \times 5$  & 593    &593&593&0.06   &  & 593   &593    &593  & -    & -  & 593 \\
 LA06   & $15 \times 5$  & 926    &926&926&0.09   &  & 926   &926    &926  & -    & -  & 926 \\
 LA07   & $15 \times 5$  & 890    &890&890&0.06   &  & 890   & 890    &890  & -    & -  & 890 \\
 LA08   & $15 \times 5$  & 863    &863&863&0.08   &  & 863   &863    &863  & -    & -  & 863 \\
 LA09   & $15 \times 5$  & 951  &951  &951&0.09   &  & 951   &951    &951  & -    & -  & 951 \\
 LA10   & $15 \times 5$  & 958  &958  &958&0.09   &  & 958   &958    &958  & -    & -  & 958 \\
 LA11   & $20 \times 5$  & 1222 &1222 &1222&0.11  &  & 1222  &1222   &1222 & -    & -  & 1222 \\
 LA12   & $20 \times 5$  & 1039 &1039 &1039&0.12  &  & 1039  &1039   &1039 & -    & -  & 1039 \\
 LA13   & $20 \times 5$  & 1150 &1150 &1150&0.12  &  & 1150  &1150   &1150 & -    & -  & 1150 \\
 LA14   & $20 \times 5$  & 1292 &1292 &1292&0.12  &  & 1292  &1292   &1292 & -    & -  & 1292 \\
 LA15   & $20 \times 5$  & 1207 &1207 &1207&0.11  &  & 1207  &1207   &1207 & -    & -  & 1207 \\
 LA16   & $10 \times 10$ & 945  &945  &945&0.15   &  & 945   &945    &945  & -    & -  & 945 \\
 LA17   & $10 \times 10$ & 784  &784  &784&0.08   &  & 784   &784    &784  & -    & -  & 784 \\
 LA18   & $10 \times 10$ & 848  &848  &848&0.09   &  & 848   &848    &848  & -    & -  & 848 \\
 LA19   & $10 \times 10$ & 842  &842  &842&0.16   &  & 842   &842    &842  &842   &842 & 844 \\
 LA20   & $10 \times 10$ & 902  &902  &902&0.11   &  & 902   &902     &902  & -    & -  & 907 \\
 LA21   & $15 \times 10$ & 1046 &1046 &1046&7.33  &  & 1046  &1046    &1046 &1046  &1046& 1046 \\
 LA22   & $15 \times 10$ & 927  &927  &927&3.94   &  & 927   &927     &927  & -    & -  & 935 \\
 LA23   & $15 \times 10$ & 1032 &1032 &1032&0.13  &  & 1032  &1032    &1032 & -    & -  & 1032 \\
 LA24   & $15 \times 10$ & 935  &935  &935&3.09   &  & 935   &935     &935  &935   &935 & 953 \\
 LA25   & $15 \times 10$ & 977  &977  &977&1.38   &  & 977   &977     &977  &977   &977 & 981 \\
 LA26   & $20 \times 10$ & 1218 &1218 &1218&0.28  &  & 1218  &1218    &1218 &-     & -  & 1218 \\
 LA27   & $20 \times 10$ & 1235 &1235 &1235&2.19  &  & 1235  &1235    &1235 &1235  &1235& 1236 \\
 LA28   & $20 \times 10$ & 1216 &1216 &1216&0.35  &  & 1216  &1216    &1216 & -    & -  & 1216 \\
 LA29   & $20 \times 10$ & 1152 &1153 &1153&73.78 &  & 1153  &1154.7  &1152 &1153  &1156& 1160 \\
 LA30   & $20 \times 10$ & 1355 &1355 &1355&0.31  &  & 1355  &1355    &1355 & -    & -  & 1355 \\
 LA31   & $30 \times 10$ & 1784 &1784 &1784&0.27  &  & 1784  &1784    &1784 & -    & -  & 1784 \\
 LA32   & $30 \times 10$ & 1850 &1850 &1850&0.29  &  & 1850  &1850    &1850 & -    & -  & 1850 \\
 LA33   & $30 \times 10$ & 1719 &1719 &1719&0.27  &  & 1719  &1719    &1719 & -    & -  & 1719 \\
 LA34   & $30 \times 10$ & 1721 &1721 &1721&0.28  &  & 1721  &1721    &1721 & -    & -  & 1721 \\
 LA35   & $30 \times 10$ & 1888 &1888 &1888&0.27  &  & 1888  &1888    &1888 & -    & -  & 1888 \\
 LA36   & $15 \times 15$ & 1268 &1268 &1268&4.53  &  & 1268  &1268    &1268 &1268  &1268& 1287 \\
 LA37   & $15 \times 15$ & 1397 &1397 &1397&26.24 &  & 1397  &1397    &1397 &1397  &1397& 1407 \\
 LA38   & $15 \times 15$ & 1196 &1196 &1196&32.61 &  & 1196  &1196    &1196 &1196  &1196& 1196 \\
 LA39   & $15 \times 15$ & 1233 &1233 &1233&11.63 &  & 1233  &1233    &1233 &1233  &1233& 1233 \\
 LA40   & $15 \times 15$ & 1222 &1222 &1222&384.8 &  & 1222  &1223.2  &1222 &1224  &1224& 1229 \\
 \hline
 MRE    &                & 0    &0.002&    &13.90 &  & 0.002 &        &0.000&0.023 &0.046& 0.189 \\
 TS/PR  &                &      &     &    &      &  & 0.002 &        &0.002&0.008 &0.008& 0.002 \\
\hline
\end{tabular}
}
\end{scriptsize}
\end{table}

\subsection{Computational results on the third set of instances}
\label{subsec Computational Results 2}

\begin{table}[!h]
\begin{scriptsize}
\caption{Computational results and comparisons for the third set of instances ABZ07-09 and YN01-04}
\label{table abz yn}
\scalebox{0.8}{
\begin{tabular}
{p{1.1 cm}p{1.3 cm}p{1.3 cm} p{1cm}p{1cm}p{1cm} p{0.01cm} p{1cm}  p{1cm} p{1cm}p{1cm} p{0.8cm} }
\hline
 & & & \multicolumn{3}{c}{TS/PR}& & \multicolumn{2}{c}{BRKGA} & & &     \\
\cline{4-6} \cline{8-9}
 \centering{\raisebox{2.50ex}[0cm][0cm]{Problem}} & \centering{\raisebox{2.50ex}[0cm][0cm]{Size}} &
 \raisebox{2.50ex}[0cm][0cm]{UB(LB)}  & Best & $M_{av}$ & $T_{av}$  & & Best & $M_{av}$ &\raisebox{2.50ex}[0cm][0cm]{GES/TS} &\raisebox{2.50ex}[0cm][0cm]{TS/SA}
 &\raisebox{2.50ex}[0cm][0cm]{TS}
    \\  \hline
ABZ07 & $20 \times 15$ &656(656) &657 &657.1&438.01 & &656 &658  &658 &658 &657\\
ABZ08 & $20 \times 15$ &665(645) &667 &667.8&138.97 & &667 &667.7&669 &669 &669\\
ABZ09 & $20 \times 15$ &678(661) &678 &678  &90.22  & &678 &678.9&679 &678 &680\\
YN01  & $20 \times 20$ &884(826) &884 &885.5&169.29 & &884 &886  &884 &884 &-  \\
YN02  & $20 \times 20$ &904(861) &904 &907.7&202.22 & &904 &906.5&905 &907 &-  \\
YN03  & $20 \times 20$ &892(827) &892 &893.8&344.15 & &892 &893.1&892 &892 &-  \\
YN04  & $20 \times 20$ &968(918) &968 &969.1&320.51 & &968 &973  &969 &969 &-  \\
\hline
MRE   &                &         &4.494&    &243.34 & &4.472& &4.614&4.625& 2.249  \\
TS/PR &                &         &    &     &       & &4.494& &4.494&4.494&2.045 \\
\hline
\end{tabular}
}
\end{scriptsize}
\end{table}

In order to further evaluate the performance of TS/PR, we tested it on the third set of benchmark JSP instances ABZ07-09 and YN01-04

Table \ref{table abz yn} summarizes the results of this experiment. In this table, the column UB(LB) lists the best known upper bound (lower bound) and the next three columns Best, M$_{av}$, and T$_{av}$ show the best makespan, average makespan, and average computing time in seconds to obtain the best value, respectively, by TS/PR over ten independent runs. The next two columns present the best makespan and average makespan of BRKGA. The last three columns show the best results of GES/TS, TS/SA, and TS, respectively. From Table \ref{table abz yn}, we see that TS/PR outperforms GES/TS, TS/SA, and TS, while it is only slightly worse than BRKGA in terms of solution quality.

\subsection{Computational results on the fourth set of instances}
\label{subsec results 3}

\begin{table}[!h]
\begin{scriptsize}
\caption{Computational results and comparisons for the fourth set of instances SWV01-15}
\label{table swv}
\scalebox{0.8}{
\begin{tabular}
{p{1.2cm}p{1.3cm}p{1.3cm} p{1cm}p{1cm}p{1cm} p{0.01cm}p{1cm} p{1cm} p{1cm}p{1cm} p{0.8cm} }
\hline
 & & & \multicolumn{3}{c}{TS/PR} &&\multicolumn{2}{c}{BRKGA}  & & &    \\
\cline{4-6}\cline{8-9}
 \centering{\raisebox{2.50ex}[0cm][0cm]{Problem}} & \centering{\raisebox{2.50ex}[0cm][0cm]{Size}} &
 \centering{\raisebox{2.50ex}[0cm][0cm]{UB(LB)}}  & Best  & $M_{av}$ & $T_{av}$  & & Best  & $M_{av}$  &\centering{\raisebox{2.50ex}[0cm][0cm]{GES/TS}} &\raisebox{2.50ex}[0cm][0cm]{TS/SA}
 &\raisebox{2.50ex}[0cm][0cm]{TS}
    \\  \hline
 SWV01 & $20 \times 10$ &1407(1407) &1407 &1411.4&575.76           & &1407&1408.9 &1412 &1412 & - \\
 SWV02 & $20 \times 10$ &1475(1475) &1475 &1475.1&294.13           & &1475&1478.2 &1475 &1475 & - \\
 SWV03 & $20 \times 10$ &1398(1369) &1398 &1398.9&613.00           & &1398&1400 &1398 &1398 & - \\
 SWV04 & $20 \times 10$ &1470(1450) &1470 &1473.5&257.63           & &1470&1472.8 &1471 &1470 & - \\
 SWV05 & $20 \times 10$ &1424(1424) &1425 &1426&612.78           & &1425&1431.4 &1426 &1425 & - \\
 SWV06 & $20 \times 15$ &1672(1591) &\textbf{1671}&1675.9&385.73   & &1675&1682.1 &1677 &1679 & - \\
 SWV07 & $20 \times 15$ &1594(1446) &1595 &1605&626.46           & &1594&1601.2 &1595 &1603 & - \\
 SWV08 & $20 \times 15$ &1752(1640) &1752 &1760.4&503.00           & &1755&1764.3 &1766 &1756 & - \\
 SWV09 & $20 \times 15$ &1656(1604) &\textbf{1655}&1661.8&521.91   & &1656&1667.9  &1660 &1661 & - \\
 SWV10 & $20 \times 15$ &1743(1631) &1743&1756.6&441.40            & &1743&1754.6  &1760 &1754 & - \\
 SWV11 & $50 \times 10$ &2983(2983) &2983 &2984.5&940.68           & &2983&2985.9  &- &-    &2983 \\
 SWV12 & $50 \times 10$ &2979(2972) &\textbf{2977}&2985.3&6097.35  & &2979&2989.7  &- &-    &2979 \\
 SWV13 & $50 \times 10$ &3104(3104) &3104 &3104&1111.22          & &3104&3111.6  &- &-    &3104 \\
 SWV14 & $50 \times 10$ &2968(2968) &2968 &2968&422.81           & &2968&2968  &- &-    &2968 \\
 SWV15 & $50 \times 10$ &2886(2885) &$\textbf{2885}^{*}$&2889.4&6000.57&&2901&2902.9  &- &- &2886 \\
 \hline
 MRE   &                &           &2.396 &      &1293.63         & &2.466&&3.886&3.848&0.054   \\
 TS/PR &                &           &      &      &                & &2.396&&3.578&3.578&0.034   \\
\hline
\end{tabular}
}
\flushleft{Newly found upper bounds by TS/PR are indicated in bold. \\
$*$: the best solution found by the TS/PR is equal to the lower bound.}
\end{scriptsize}
\end{table}

The set of benchmark JSP instances SWV01-15 was first reported by \cite{Storer1992SWV01-15}. As this set contains some of the most difficult JSP instances, many powerful algorithms have been proposed for solving them. However, 60\% of these instances have not been solved until now. 

From Table \ref{table swv}, it is easy to see that TS/PR outperforms all of the five reference algorithms in terms of the MRE value. Specifically, TS/PR matches nine best known solutions and improves the upper bounds for the four instances SWV06, SWV09, SWV12, and SWV15. Even if TS/PR cannot reach the best upper bound for the two instances SWV05 and SWV07, the gaps between our solutions and the best upper bounds are only one unit. In comparison, BRKGA, which is one of the best performing algorithms in the literature, reaches the upper bounds for 11 out of 15 instances. It is worth noting that the best makespan obtained by TS/PR is better than that of BRKGA in four cases, while TS/PR is worse than BRKGA for one instance.

In particular, TS/PR is able to find better upper bounds for the instances SWV06-1671, SWV09-1655, SWV12-2977, and SWV15-2885, while the previous best upper bounds were for the instances SWV06-1672, SWV09-1656, SWV12-2979, and SWV15-2886. More strikingly, TS/PR is able to reach the best lower bound for the instance SWV15, meaning that TS/PR solves this instance, which has remained unsolved for over 20 years.

\subsection{Computational results on the fifth set of instances}
\label{subsec results 4}

The fifth set of 50 benchmark JSP instances TA is one of the most widely used set of instances and is also part of the most difficult JSP instances over the last 20 years.

Table \ref{table ta} presents the results of applying TS/PR to tackle the set of TA instances and comparisons with the reference algorithms. As can be seen, TS/PR is able to find the current best known solutions for 34 of the 50 instances and in addition improve upon the solutions for five instances. Specifically, TS/PR finds better upper bounds for the instances TA43-1846, TA44-1982, TA47-1889, TA49-1963, and TA50-1923, while the previous best upper bounds were for the instances TA43-1848, TA44-1983, TA47-1894, TA49-1964, and TA50-1924. Although TS/PR performs slightly worse than BRKGA in terms of the MRE value, it outperforms all the other reference algorithms in the literature.

\begin{table}[!hp]
\begin{scriptsize}
\caption{Computational results and comparisons for the fifth set of instances TA01-50}
\label{table ta}
\scalebox{0.64}{
\begin{tabular}
{p{1.1cm}p{1.5cm}p{1.5cm} p{1.1cm}p{1.1cm}p{1.1cm}  p{0.01cm} p{1.1cm} p{1.1cm} p{1.1cm}p{1.1cm} p{1.1cm}p{1.1cm}p{1.1cm} }
\hline
 & & & \multicolumn{3}{c}{TS/PR} &&\multicolumn{2}{c}{BRKGA} & & & & &    \\
\cline{4-6}\cline{8-9}
 \centering{\raisebox{2.50ex}[0cm][0cm]{Problem}} & \centering{\raisebox{2.50ex}[0cm][0cm]{Size}} &
 \centering{\raisebox{2.50ex}[0cm][0cm]{UB(LB)}}  & Best & $M_{av}$ & $T_{av}$ &
 & Best & $M_{av}$ & \raisebox{2.50ex}[0cm][0cm]{CP/LS}  &
 \raisebox{2.50ex}[0cm][0cm]{GES}     &\centering{\raisebox{2.50ex}[0cm][0cm]{AlgFix}}&
 \raisebox{2.50ex}[0cm][0cm]{i-TSAB}  &\raisebox{2.50ex}[0cm][0cm]{TS/SA}
     \\ \hline
TA01 & $15 \times 15$ & 1231(1231) &1231 &1231&2.93     &&1231&1231 & - &1231 &1231 &  -   &1231 \\
TA02 & $15 \times 15$ & 1244(1244) &1244 &1244&38.09    &&1244&1244 & - &1244 &1244 &  -   &1244 \\
TA03 & $15 \times 15$ & 1218(1218) &1218 &1218&43.66    &&1218&1218 & - &1218 &1218 &  -   &1218 \\
TA04 & $15 \times 15$ & 1175(1175) &1175 &1175&38.72    &&1175&1175 & - &1175 &1175 &  -   &1175 \\
TA05 & $15 \times 15$ & 1224(1224) &1224 &1224&11.24    &&1224&1224.9 & - &1224 &1224 &  -   &1224 \\
TA06 & $15 \times 15$ & 1238(1238) &1238 &1238.4&178.06 &&1238&1238.9 & - &1238 &1238 &  -   &1238 \\
TA07 & $15 \times 15$ & 1227(1227) &1228 &1228&0.60     &&1228&1228 & - &1228 &1228 &  -   &1228 \\
TA08 & $15 \times 15$ & 1217(1217) &1217 &1217&2.43     &&1217&1217 & - &1217 &1217 &  -   &1217 \\
TA09 & $15 \times 15$ & 1274(1274) &1274 &1274&18.66    &&1274&1277 & - &1274 &1274 &  -   &1274 \\
TA10 & $15 \times 15$ & 1241(1241) &1241 &1241&42.25    &&1241&1241 & - &1241 &1241 &  -   &1241 \\
TA11 & $20 \times 15$ & 1357(1323) &1357 &1359.9&186.19 &&1357&1360 &1357 &1357 &1358 &1361 &1359 \\
TA12 & $20 \times 15$ & 1367(1351) &1367 &1369.9&206.06 &&1367&1372.6 &1367 &1367 &1367 &  -  &1371 \\
TA13 & $20 \times 15$ & 1342(1282) &1342 &1346&161.37   &&1344&1347.3 &1342 &1344 &1342 &  -  &1342 \\
TA14 & $20 \times 15$ & 1345(1345) &1345 &1345&8.28     &&1345&1345 &1345 &1345 &1345 &  -  &1345 \\
TA15 & $20 \times 15$ & 1339(1304) &1339 &1339&173.45   &&1339&1348.9 &1339 &1339 &1339 &  -  &1339 \\
TA16 & $20 \times 15$ & 1360(1304) &1360 &1360&63.41    &&1360&1362.1 &1360 &1360 &1360 &  -  &1360 \\
TA17 & $20 \times 15$ & 1462(1462) &1463&1473&203.49    &&1462&1470.5 &1462 &1469 &1473 &1462 &1464 \\
TA18 & $20 \times 15$ & 1396(1369) &1396 &1401&91.13    &&1396&1400.9 &1396 &1401 &1396 &  -  &1399 \\
TA19 & $20 \times 15$ & 1332(1304) &1332 &1336.6&145.42 &&1332&1333.2 &1332 &1332 &1332 &1335 &1335 \\
TA20 & $20 \times 15$ & 1348(1318) &1348 &1351.3&216.72 &&1348&1350.4 &1348 &1348 &1348 &1351 &1350 \\
TA21 & $20 \times 20$ & 1642(1573) &1644&1645.2&502.99  &&1642&1647 &1642 &1647 &1643 &1644 &1644 \\
TA22 & $20 \times 20$ & 1600(1542) &1600 &1603.8&228.90 &&1600&1600 &1600 &1602 &1600 &1600 &1600 \\
TA23 & $20 \times 20$ & 1557(1474) &1557 &1559.6&359.79 &&1557&1562.6 &1557 &1558 &1557 &1557 &1560 \\
TA24 & $20 \times 20$ & 1644(1606) &1645&1647.7&779.32  &&1646&1650.6 &1644 &1653 &1646 &1647 &1646 \\
TA25 & $20 \times 20$ & 1595(1518) &1595 &1597&416.08   &&1595&1602 &1595 &1596 &1595 &1595 &1597 \\
TA26 & $20 \times 20$ & 1643(1558) &1647 &1651.4&267.50 &&1643&1652.3 &1643 &1647 &1647 &1645 &1647 \\
TA27 & $20 \times 20$ & 1680(1617) &1680 &1686.7&254.74 &&1680&1685.6 &1680 &1685 &1686 &1680 &1680 \\
TA28 & $20 \times 20$ & 1603(1591) &1613 &1616.2&326.23 &&1603&1611.7 &1603 &1614 &1613 &1614 &1603 \\
TA29 & $20 \times 20$ & 1625(1525) &1625 &1627.4&93.53  &&1625&1627.4 &1625 &1625 &1625 &  -  &1627 \\
TA30 & $20 \times 20$ & 1584(1485) &1584 &1588.3&388.66 &&1584&1588.5 &1584 &1584 &1584 &1584 &1584 \\
TA31 & $30 \times 15$ & 1764(1764) &1764 &1764&35.57    &&1764&1764.4 &1764 &1764 &1766 &  -  &1764 \\
TA32 & $30 \times 15$ & 1785(1774) &1787 &1803.5&703.06 &&1785&1794.1 &1796 &1793 &1790 &  -  &1795 \\
TA33 & $30 \times 15$ & 1791(1778) &1791 &1794.6&457.55 &&1791&1793.7 &1791 &1799 &1791 &1793 &1796 \\
TA34 & $30 \times 15$ & 1829(1828) &1829 &1831.2&315.71 &&1829&1832.1 &1829 &1832 &1832 &1829 &1831 \\
TA35 & $30 \times 15$ & 2007(2007) &2007 &2007&0.56 &&2007&2007&2007 &2007 &2007 &  -  &2007 \\
TA36 & $30 \times 15$ & 1819(1819) &1819 &1819&122.67   &&1819&1822.9 &1819 &1819 &1819 &  -  &1819 \\
TA37 & $30 \times 15$ & 1771(1771) &1771 &1776.8&652.24 &&1771&1777.8 &1774 &1779 &1784 &1778 &1778 \\
TA38 & $30 \times 15$ & 1673(1673) &1673 &1673&307.34   &&1673&1676.7 &1673 &1673 &1673 &  -  &1673 \\
TA39 & $30 \times 15$ & 1795(1795) &1795 &1795&115.61   &&1795&1801.6 &1795 &1795 &1795 &  -  &1795 \\
TA40 & $30 \times 15$ & 1669(1631) &1671&1676&449.96    &&1669&1678.1 &1673 &1680 &1979 &1674 &1676 \\
TA41 & $30 \times 20$ & 2006(1874) &2010&2018.6&1267.78 &&2008&2018.7 &2010 &2022 &2022 &  -  &2018 \\
TA42 & $30 \times 20$ & 1937(1867) &1949&1950.3&1556.36             &&1937&1949.3 &1947 &1956 &1953 &1956 &1953 \\
TA43 & $30 \times 20$ & 1848(1809) &\textbf{1846}&1865.1&1726.78    &&1852&1863.1 &1863 &1870 &1869 &1859 &1858 \\
TA44 & $30 \times 20$ & 1983(1927) &\textbf{1982}&1989.1&1304.66    &&1983&1992.4 &1991 &1991 &1992 &1984 &1983 \\
TA45 & $30 \times 20$ & 2000(1997) &2000 &2000.5&1057.79            &&2000&2000 &2000 &2004 &2000 &2000 &2000 \\
TA46 & $30 \times 20$ & 2004(1940) &2008&2022.3&1236.03             &&2004&2015.5 &2016 &2011 &2011 &2021 &2010 \\
TA47 & $30 \times 20$ & 1894(1789) &\textbf{1889}&1906.2&1030.88    &&1894&1902.1 &1906 &1903 &1902 &1903 &1903 \\
TA48 & $30 \times 20$ & 1943(1912) &1947&1955.5&1047.42             &&1943&1959.2 &1951 &1962 &1962 &1953 &1955  \\
TA49 & $30 \times 20$ & 1964(1915) &\textbf{1963}&1971.5&1035.82    &&1964&1972.6 &1966 &1969 &1974 &-    &1967 \\
TA50 & $30 \times 20$ & 1924(1807) &\textbf{1923}&1931.4&1318.05    &&1925&1927 &1924 &1931 &1927 &1928 &1931 \\
\hline
MRE  &                 &            &2.162         &     &423.83&&2.133 & &2.769&2.356&2.688&3.233&2.279\\
TS/PR&                 &            &              &     &      &&2.162 & &2.701&2.162&2.162&3.046&2.162\\
\hline
\end{tabular}
}
\flushleft{Newly found upper bounds by TS/PR are indicated in bold.}
\end{scriptsize}
\end{table}




\subsection{Computational results on the sixth set of instances}
\label{subsec results 5}

Our last experiment was based on the DMU set of instances, which are considered to be one of the hardest JSP instances. In particular, the instances DMU41-80 are considered to be extremely challenging \citep{Demirkol1997DMU01-80}. However, our computational experiments show that TS/PR yields high-quality solutions for these instances and can even improve many of the best upper bounds. The detailed results for these instances are presented in Tables \ref{table dmu01} and \ref{table dmu41}.

In general, TS/PR performs well on these DMU instances in comparison with the reference algorithms BRKGA, TS, GES, i-TSAB, and AlgFix. The results reveal that TS/PR outperforms all of these algorithms for the majority of these instances. In particular, for the first 40 DMU instances, TS/PR is able to improve the best upper bounds for five instances and solve the problem in less CPU time than BRKGA, one of the best performing algorithms for these instances. 

On the other hand, TS/PR is able to obtain better solutions for the difficult instances DMU41-80. For example, TS/PR is able to improve 35 upper bounds and hit two best upper bounds for these instances.

In sum, TS/PR finds improved upper bounds for 40 out of the 80 DMU instances, i.e., 50\% of this set of the hardest JSP instances. This experiment demonstrates the competitiveness of TS/PR in terms of both solution quality and computational efficiency.

\begin{table}[hp]
\begin{scriptsize}
\caption{Computational results and comparisons for the sixth set of instances DMU01-40}
\label{table dmu01}
\scalebox{0.8}{
\begin{tabular}
{p{1cm}p{1cm}p{1.5cm} p{0.9cm}p{0.9cm}p{0.9cm} p{0.1cm}p{0.9cm} p{0.9cm} p{0.9cm}p{0.9cm} p{0.9cm}p{0.9cm} }
\hline
 & & & \multicolumn{3}{c}{TS/PR} && \multicolumn{2}{c}{BRKGA} & & & &     \\
\cline{4-6}\cline{8-9}
 \centering{\raisebox{2.50ex}[0cm][0cm]{Problem}} & \centering{\raisebox{2.50ex}[0cm][0cm]{Size}} &
 \centering{\raisebox{2.50ex}[0cm][0cm]{UB(LB)}}  & Best  & $M_{av}$ & $T_{av}$            & & Best  & $M_{av}$
  & \raisebox{2.50ex}[0cm][0cm]{TS}  &
 \raisebox{2.50ex}[0cm][0cm]{GES}     &\centering{\raisebox{2.50ex}[0cm][0cm]{i-TSAB}}&
 \raisebox{2.50ex}[0cm][0cm]{AlgFix}
     \\ \hline
 DMU01   & $20 \times 15$ & 2563(2501) &2563&2563&332.87      && 2563&2563    & 2566 & 2566 & 2571  & 2563  \\
 DMU02   & $20 \times 15$ & 2706(2651) &2706&2713.2&179.24    && 2706&2714.5    & 2711 & 2706 & 2715  & 2706  \\
 DMU03   & $20 \times 15$ & 2731(2731) &2731&2733.1&388.59    && 2731&2736.5    &  -   & 2731 &  -    & 2731  \\
 DMU04   & $20 \times 15$ & 2669(2601) &2669&2670.2&96.54     && 2669&2672.4    &  -   & 2669 &  -    & 2669  \\
 DMU05   & $20 \times 15$ & 2749(2749) &2749&2758.6&303       && 2749&2755.4    &  -   & 2749 &   -   & 2749  \\
 DMU06   & $20 \times 20$ & 3244(2998) &3245&3249.2 &823.17   && 3244&3246.6    & 3254 & 3250 & 3265  & 3244  \\
 DMU07   & $20 \times 20$ & 3046(2815) &3046&3062.3 &360.58   && 3046&3058.6    &  -   & 3053 &  -    & 3046  \\
 DMU08   & $20 \times 20$ & 3188(3051) &3188&3194.3 &295.81   && 3188&3188.3    & 3191 & 3197 & 3199  & 3188  \\
 DMU09   & $20 \times 20$ & 3092(2956) &3094&3097.4 &148.00   && 3092&3094.4    &  -  & 3092 & 3094  & 3096  \\
 DMU10   & $20 \times 20$ & 2984(2858) &2985&2991 &252.46   && 2984&2984.8    &  -   & 2984 & 2985  & 2984  \\
 DMU11   & $30 \times 15$ & 3445(3395) &\textbf{3430}&3435.2 &1496.85 && 3445&3445.8    & 3455 & 3453 & 3470  & 3455 \\
 DMU12   & $30 \times 15$ & 3513(3481) &\textbf{3495}&3509.7 &899.99  && 3513&3518.9    & 3516 & 3518 & 3519  & 3522 \\
 DMU13   & $30 \times 15$ & 3681(3681) &3681&3682.8 &622.13   && 3681&3690.6    & 3681 & 3697 & 3698  & 3687  \\
 DMU14   & $30 \times 15$ & 3394(3394) &3394&3394&3.02        && 3394&3394 &  -   & 3394 & 3394  & 3394  \\
 DMU15   & $30 \times 15$ & 3343(3343) &3343&3343&1.77        && 3343&3343 &  -   & 3343 &   -   & 3343  \\
 DMU16   & $30 \times 20$ & 3751(3734) &3753&3765.4 &1303.41  && 3751&3758.9    & 3759 & 3781  & 3787  & 3772  \\
 DMU17   & $30 \times 20$ & 3830(3709) &\textbf{3819}&3843.3 &734.03 &&3830& 3850.6    & 3842 & 3848 & 3854  & 3836 \\
 DMU18   & $30 \times 20$ & 3844(3844) &3844&3849.5 &3787.40  && 3844&3845.4    & 3846 & 3849 & 3854  & 3852  \\
 DMU19   & $30 \times 20$ & 3770(3669) &\textbf{3768}&3787.4 &718.71 &&3770&3791.8    & 3784 & 3807 & 3823  & 3775 \\
 DMU20   & $30 \times 20$ & 3712(3604) &\textbf{3710}&3726.5 &701.29 &&3712&3715.3    & 3716 & 3739 & 3740  & 3712  \\
 DMU21   & $40 \times 15$ & 4380(4380)  &4380&4380&0.69      && 4380&4380    &  -   & 4380 &   -   & 4380  \\
 DMU22   & $40 \times 15$ & 4725(4725)  &4725&4725&1.48      && 4725&4725    &  -   & 4725 &   -   & 4725  \\
 DMU23   & $40 \times 15$ & 4668(4668)  &4668&4668&1.30      && 4668&4668    &  -   & 4668 &   -   & 4668  \\
 DMU24   & $40 \times 15$ & 4648(4648)  &4648&4648&0.75      && 4648&4648    &  -   & 4648 &   -   &4648  \\
 DMU25   & $40 \times 15$ & 4164(4164)  &4164&4164&0.60      && 4164&4164    &  -   & 4164 &   -   &4164  \\
 DMU26   & $40 \times 20$ & 4647(4647)  &4647&4647.3&1631.43 && 4647&4658.4    & 4647 & 4667 & 4679  &4688  \\
 DMU27   & $40 \times 20$ & 4848(4848)  &4848&4848&12.16     && 4848&4848    &  -  & 4848 & 4848  &4848  \\
 DMU28   & $40 \times 20$ & 4692(4692)  &4692&4692&17.68     && 4692&4692    &  -   & 4692 &  -    &4692  \\
 DMU29   & $40 \times 20$ & 4691(4691)  &4691&4691&63.49     && 4691&4691    &  -   & 4691 & 4691  &4691  \\
 DMU30   & $40 \times 20$ & 4732(4732)  &4732&4732&123.00    && 4732&4732    &  -   & 4732 & 4732  &4749  \\
 DMU31   & $50 \times 15$ & 5640(5640)  &5640&5640&0.84      && 5640&5640    &  -   & 5640 &   -   &5640  \\
 DMU32   & $50 \times 15$ & 5927(5927)  &5927&5927&0.62      && 5927&5927    &  -   & 5927 &   -   &5927  \\
 DMU33   & $50 \times 15$ & 5728(5728)  &5728&5728&0.43      && 5728&5728    &  -   & 5728 &   -   &5728  \\
 DMU34   & $50 \times 15$ & 5385(5385)  &5385&5385&2.22      && 5385&5385    &  -   & 5385 &   -   &5385  \\
 DMU35   & $50 \times 15$ & 5635(5635)  &5635&5635&0.71      && 5635&5635    &  -   & 5635 &   -   &5635  \\
 DMU36   & $50 \times 20$ & 5621(5621)  &5621&5621&7.83      && 5621&5621    &  -   & 5621 &   -   &5621  \\
 DMU37   & $50 \times 20$ & 5851(5851)  &5851&5851&11.38     && 5851&5851    &  -   & 5851 & 5851  &5851  \\
 DMU38   & $50 \times 20$ & 5713(5713)  &5713&5713&10.66     && 5713&5713    &  -   & 5713 &   -   &5713  \\
 DMU39   & $50 \times 20$ & 5747(5747)  &5747&5747&2.02      && 5747&5747    &  -   & 5747 &   -   &5747  \\
 DMU40   & $50 \times 20$ & 5577(5577)  &5577&5577&4.91      && 5577&5577    &  -   & 5577 &  -    &5577  \\
\hline
\end{tabular}
}
\flushleft{Newly found upper bounds by TS/PR are indicated in bold.}
\end{scriptsize}
\end{table}

\begin{table}[hp]
\begin{scriptsize}
\caption{Computational results and comparisons for the sixth set of instances DMU41-80}
\label{table dmu41}
\scalebox{0.78}{
\begin{tabular}
{p{1cm}p{1cm}p{1.5cm} p{0.9cm}p{0.9cm}p{0.9cm} p{0.1cm}p{0.9cm} p{0.9cm} p{0.9cm}p{0.9cm} p{0.9cm}p{0.9cm} }
\hline
 & & & \multicolumn{3}{c}{TS/PR} && \multicolumn{2}{c}{BRKGA} & & & &     \\
\cline{4-6}\cline{8-9}
 \centering{\raisebox{2.50ex}[0cm][0cm]{Problem}} & \centering{\raisebox{2.50ex}[0cm][0cm]{Size}} &
 \centering{\raisebox{2.50ex}[0cm][0cm]{UB(LB)}}  & Best  & $M_{av}$ & $T_{av}$            & & Best  & $M_{av}$
  & \raisebox{2.50ex}[0cm][0cm]{TS}  &
 \raisebox{2.50ex}[0cm][0cm]{GES}     &\centering{\raisebox{2.50ex}[0cm][0cm]{i-TSAB}}&
 \raisebox{2.50ex}[0cm][0cm]{AlgFix}
     \\ \hline
DMU41 & $20 \times 15$ & 3261(3007) &\textbf{3248} &3281.9&417.84    &&3261&3281.9 & -   &3267 &3277 &3278 \\
DMU42 & $20 \times 15$ & 3395(3172) &\textbf{3390} &3409.8&448.95    &&3395&3403.9 &3416 &3401 &3448 &3412 \\
DMU43 & $20 \times 15$ & 3441(3292) &3441 &3450.5 &399.33            &&3441&3452.7 &3459 &3443 &3473 &3450 \\
DMU44 & $20 \times 15$ & 3488(3283) &3489 &3509.7 &371.27            &&3488&3510.7 &3524 &3489 &3528 &3489 \\
DMU45 & $20 \times 15$ & 3272(3001) &3273 &3287.9 &709.19            &&3272&3287.3 &3296 &3273 &3321 &3273 \\
DMU46 & $20 \times 20$ & 4035(3575) &4035 &4051.8 &984.86            &&4035&4043.2 &4080 &4099 &4101 &4071 \\
DMU47 & $20 \times 20$ & 3939(3522) &3942 &3963.6 &829.28            &&3939&3968 & - &3972 &3973 &3950 \\
DMU48 & $20 \times 20$ & 3781(3447) &\textbf{3778} &3814.1 &938.55   &&3781&3800.9 &3795 &3810 &3838 &3813 \\
DMU49 & $20 \times 20$ & 3723(3403) &\textbf{3710} &3736.1&633.84    &&3723&3729.6 &3735 &3754 &3780 &3725 \\
DMU50 & $20 \times 20$ & 3732(3496) &\textbf{3729} &3741.2 &609.62   &&3732&3746.5 &3761 &3768 &3794 &3742 \\
DMU51 & $30 \times 15$ & 4201(3917) &\textbf{4167} &4205.9 &2394.25  &&4201&4222.9 &4218 &4247 &4260 &4202 \\
DMU52 & $30 \times 15$ & 4341(4065) &\textbf{4311} &4353.2 &2232.60  &&4341&4352.3 &4362 &4380 &4383 &4353 \\
DMU53 & $30 \times 15$ & 4415(4141) &\textbf{4394} &4425.7 &2161.83  &&4415&4420.2 &4428 &4450 &4470 &4419 \\
DMU54 & $30 \times 15$ & 4396(4202) &\textbf{4371} &4390.5 &1909.53  &&4396&4402.7 &4405 &4424 &4425 &4413 \\
DMU55 & $30 \times 15$ & 4290(4140) &\textbf{4271} &4295.2 &1914.37  &&4290&4299.4 &4308 &4331 &4332 &4321 \\
DMU56 & $30 \times 20$ & 4961(4554) &\textbf{4941} &4990.6 &3825.44  &&4961&4768.4 &5025 &5051 &5079 &4985 \\
DMU57 & $30 \times 20$ & 4698(4302) &\textbf{4663} &4714 &3649.41    &&4698&4704.9 &4698 &4779 &4785 &4709 \\
DMU58 & $30 \times 20$ & 4751(4319) &\textbf{4708} &4779.4 &3639.68  &&4751&4752.8 &4796 &4829 &4834 &4787 \\
DMU59 & $30 \times 20$ & 4630(4217) &\textbf{4624} &4670.6 &3614.54  &&4630&4633.3 &4667 &4694 &4696 &4638 \\
DMU60 & $30 \times 20$ & 4774(4319) &\textbf{4755} &4804.3 &3745.91  &&4774&4777 &4805 &4888 &4904 &4827 \\
DMU61 & $40 \times 15$ & 5224(4917) &\textbf{5195} &5217.1&4739.13   &&5224&5233.3 &5228 &5293 &5294 &5310 \\
DMU62 & $40 \times 15$ & 5301(5033) &\textbf{5268} &5301&4853.75     &&5301&5304.4 &5311 &5354 &5354 &5330 \\
DMU63 & $40 \times 15$ & 5357(5111) &\textbf{5326} &5347.5&4122.65   &&5357&5386.6 &5371 &5439 &5446 &5431 \\
DMU64 & $40 \times 15$ & 5312(5130) &\textbf{5252} &5279.8&4487.26   &&5312&5321.8 &5330 &5388 &5443 &5385 \\
DMU65 & $40 \times 15$ & 5197(5105) &\textbf{5196} &5203.2 &4963.8   &&5197&5211.5&5201 &5269 &5271 &5322        \\
DMU66 & $40 \times 20$ & 5796(5391) &\textbf{5717} &5788.7 &9543.86  &&5796&5806.6 &5797 &5902 &5911 &5886 \\
DMU67 & $40 \times 20$ & 5863(5589) &\textbf{5816} &5852.5 &8431.51  &&5863&5881.3 &5872 &6012 &6016 &5938 \\
DMU68 & $40 \times 20$ & 5826(5426) &\textbf{5773} &5801.8 &8739.45  &&5826&5843.7 &5834 &5934 &5936 &5840 \\
DMU69 & $40 \times 20$ & 5775(5423) &\textbf{5709} &5754.4 &8107.63  &&5775&5804 &5794 &6002 &5891 &5868 \\
DMU70 & $40 \times 20$ & 5951(5501) &\textbf{5903} &5924.2 &7285.27  &&5951&5968.2 &5954 &6072 &6096 &6028 \\
DMU71 & $50 \times 15$ & 6278(6080) &\textbf{6223} &6264.8 &9835.11  &&6293&6603.8 &6278 &6333 &6359 &6437 \\
DMU72 & $50 \times 15$ & 6503(6395) &\textbf{6483} &6510.9 &10881.79 &&6503&6560.7 &6520 &6589 &6586 &6604 \\
DMU73 & $50 \times 15$ & 6219(6001) &\textbf{6163} &6199.8 &11475.15 &&6219&6250.5 &6249 &6291 &6330 &6343 \\
DMU74 & $50 \times 15$ & 6277(6123) &\textbf{6227} &6266.4 &11164.43 &&6277&6312.6 &6316 &6376 &6383 &6467 \\
DMU75 & $50 \times 15$ & 6236(6029) &\textbf{6197} &6239.4 &11330.86 &&6248&6282.4 &6236 &6380 &6437 &6397 \\
DMU76 & $50 \times 20$ & 6876(6342) &\textbf{6813} &6854.8 &9998.17  &&6876&6885.4 &6893 &6974 &7082 &6975 \\
DMU77 & $50 \times 20$ & 6857(6499) &\textbf{6822} &6879.9 &12062.88 &&6857&6892.7 &6868 &7006 &6930 &6949 \\
DMU78 & $50 \times 20$ & 6831(6586) &\textbf{6770} &6813.2 &10346.61 &&6831&6855.7 &6846 &6988 &7027 &6928 \\
DMU79 & $50 \times 20$ & 7049(6650) &\textbf{6970} &7003   &9818.93  &&7049&7060.9 &7055 &7158 &7253 &7083 \\
DMU80 & $50 \times 20$ & 6719(6459) &\textbf{6686} &6700.1 &10331.98 &&6736&6757.9 &6719 &6843 &6998 &6861 \\
\hline
MRE   &                &            &3.596         &       &2791.17  &&3.890& &5.655& 4.669 &6.375&4.444 \\
TS/PR &                &            &              &       &         &&3.577& &4.829&3.577&4.589&3.577\\
\hline
\end{tabular}
}
\flushleft{Newly found upper bounds by TS/PR are indicated in bold.}
\end{scriptsize}
\end{table}

\subsection{Performance summary of TS/PR}
\label{subsec total performance}

\begin{figure*}[!h]
    \begin{center}
        \subfigure[The overall performance]{\includegraphics[width=0.327\textwidth]{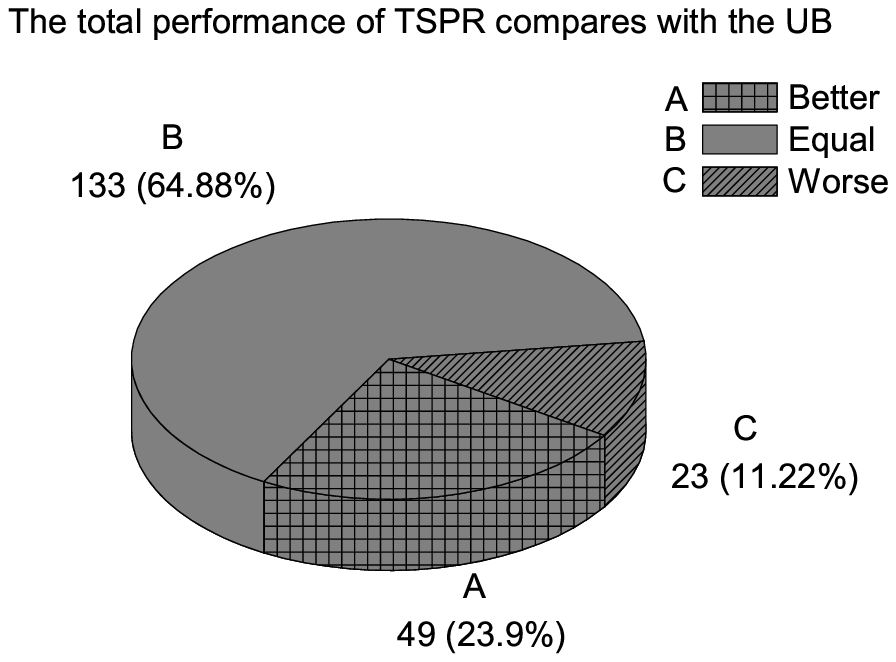}\label{total1}}
         \subfigure[Outperforming the UB]{\includegraphics[width=0.327\textwidth]{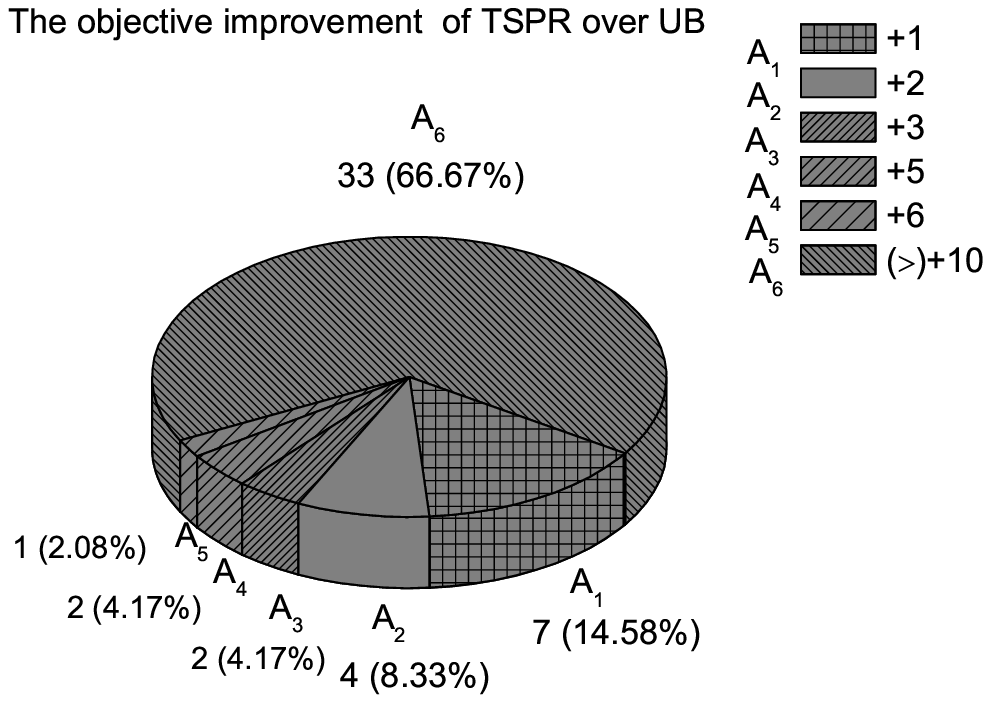}\label{total2}}
        \subfigure[Underperforming the UB]{\includegraphics[width=0.317\textwidth]{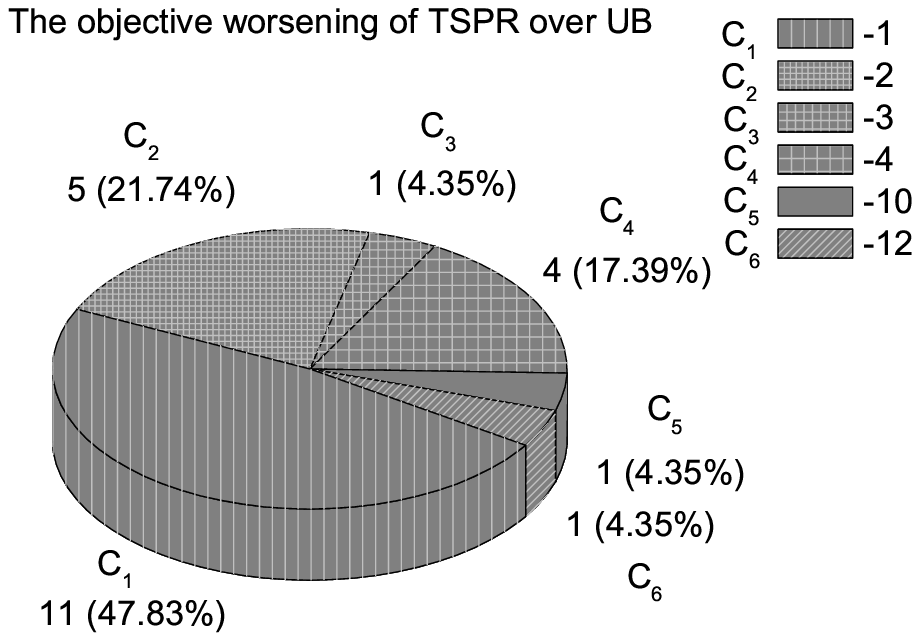}\label{total3}}
        \caption{The overall performance of TS/PR in terms of solution quality}\label{total solution quality}
    \end{center}
\end{figure*}

Finally, we summarize in Figure \ref{total solution quality} the overall performance of TS/PR in tackling all the 205 tested instances. Figure \ref{total1} presents the numbers of better, equal, and worse solutions that TS/PR is able to obtain compared with the corresponding upper bounds (UB). We see that TS/PR can improve the best upper bounds for 49 out of the 205 tested instances and tie the best upper bounds for 133 instances, while obtaining worse results for only 23 instances.

Figures \ref{total2} and \ref{total3} give the numbers of instances for which TS/PR outperforms and underperforms the UB, respectively. For example, the instance class for which TS/PR improves the UB by one unit is denoted by $A_{1}$ and it consists of seven instances, accounting for 14.58\% of the total 205 instances. It is worthwhile to note that TS/PR improves the UB by more than ten units for two thirds of the 205 instances, whereas there are only two instances for which TS/PR underperforms the UB by more than ten units. Therefore, we conclude that TS/PR not only possesses a strong improvement capability, but it also has great improvement strength.

From Table \ref{table average time}, we observe that the average computing time of TS/PR is an order of magnitude less than BRKGA for the instances DMU21-40 for which the optimal solutions are known. In particular, for the instances DMU31-35, the computing time of TS/PR is 2,000 times less than that of BRKGA. Moreover, for many of the benchmark instances for which the optimal solutions are not known, the computing times of TS/PR and BRKGA are relatively close, despite that TS/PR can usually obtain comparable or better results. In sum, TS/PR is competitive with the state-of-the-art algorithm BRKGA in terms of computational efficiency.


\label{subsec Computational Results 1}

\begin{table}[!h]
\begin{scriptsize}
\caption{Computational time comparisons with BRKGA (in seconds) for all the instances reported in this paper}
\label{table average time}
\scalebox{0.78}{
\begin{tabular}
{p{2.2cm}p{1.7cm}p{1.7cm}p{1.7cm}    p{2.2cm}p{1.7cm}p{1.7cm}p{1.2cm}}
\hline
Instance Group1  & Size    & TS/PR& BRKGA & Instance Group2  & Size    & TS/PR& BRKGA
    \\  \hline
FT06     & $6  \times 6 $ &0.03   &1.0    &TA21-30   & $20 \times 20$ &361.77    &143.2      \\
FT10     & $10 \times 10$ &4.75   &10.1   &TA31-40   & $30 \times 15$ &316.03    &487.6      \\
FT20     & $20 \times  5$ &0.18   &13.4   &TA41-50   & $30 \times 20$ &1258.16   &1068.3      \\
ORB01-10 & $10 \times 10$ &2.12   &5.8    &DMU01-05  & $20 \times 15$ &260.05    &68.9      \\
LA01-05  & $10 \times  5$ &0.05   &1.4    &DMU06-10  & $20 \times 20$ &376.00    &145.4      \\
LA06-10  & $15 \times  5$ &0.08   &2.9    &DMU11-15  & $30 \times 15$ &604.75    &427.3      \\
LA11-15  & $20 \times  5$ &0.12   &5.3    &DMU16-20  & $30 \times 20$ &1448.97   &1043.6      \\
LA16-20  & $10 \times 10$ &0.12   &4.6    &DMU21-25  & $40 \times 15$ &0.96      &1150.6      \\
LA21-25  & $15 \times 10$ &3.17   &15.3   &DMU26-30  & $40 \times 20$ &369.55    &3556.3      \\
LA26-30  & $20 \times 20$ &15.38  &21.8   &DMU31-35  & $50 \times 15$ &0.96      &2086.7      \\
LA31-35  & $30 \times 10$ &0.28   &38.7   &DMU36-40  & $50 \times 20$ &7.36      &9368.3      \\
LA36-40  & $15 \times 15$ &91.96  &21.4   &DMU41-45  & $20 \times 15$ &469.32    &78.9      \\
ABZ07-09 & $20 \times 15$ &222.40 &54.6   &DMU46-50  & $20 \times 20$ &799.23    &187.7      \\
YN01-04  & $20 \times 20$ &259.04 &105.2  &DMU51-55  & $30 \times 15$ &2122.52   &701.4      \\
SWV01-05 & $20 \times 10$ &470.66 &42.5   &DMU56-60  & $30 \times 20$ &3695.00   &1545.8      \\
SWV06-10 & $20 \times 15$ &495.70 &78.7   &DMU61-65  & $40 \times 15$ &4633.32   &2684.3      \\
SWV11-15 & $50 \times 10$ &2914.53&2304.4 &DMU66-70  & $40 \times 20$ &8421.54   &5394.2      \\
TA01-10  & $15 \times 15$ &37.66  &30.4   &DMU71-75  & $50 \times 15$ &10937.47  &8070.1      \\
TA11-20  & $20 \times 15$ &145.55 &65.8   &DMU76-80  & $50 \times 20$ &10511.71  &15923.4      \\

\hline
\end{tabular}
}
\flushleft{BRKGA was run on an AMD 2.2 GHz Opteron(2427) CPU running the Linux (Fedora release 12) operating system.}
\end{scriptsize}
\end{table}

\section{Conclusion}
\label{sec conclusions}
In this paper we present a hybrid tabu search/path relinking algorithm for tackling the notorious job shop scheduling problem, in which we incorporate a number of distinguishing features, such as a path solution construction procedure based on the distances of the solutions and a special mechanism to determine the reference solution. Based on extensive computational results of applying TS/PR to tackle six sets of a total of 205 well-known and challenging benchmark JSP instances, we demonstrate the efficacy of TS/PR in comparison with the best known results in the literature. Specifically, TS/PR is able to improve the upper bounds for 49 instances. In addition, TS/PR solves the challenging SWV15, which has remained unsolved for over 20 years.
The results confirm that the relinking method is a powerful diversification tool for tackling JSP compared with other state-of-the-art algorithms for JSP. Finally, given that many of the ideas introduced in this paper are independent of JSP, it is worthwhile to test their merits in dealing with other difficult combinatorial optimization problems.

\section*{Acknowledgment}
The research was supported in part by the Hong Kong Scholars Programme, the National Natural Science Foundation of China under grant number 61100144, 61370183, and the programme for New Century Excellent Talents in University (NCET 2013).

\bibliographystyle{elsarticle-harv}
\bibliography{mybibfile}

\begin{thebibliography}{24}
\expandafter\ifx\csname natexlab\endcsname\relax\def\natexlab#1{#1}\fi
\expandafter\ifx\csname url\endcsname\relax
  \def\url#1{\texttt{#1}}\fi
\expandafter\ifx\csname urlprefix\endcsname\relax\def\urlprefix{URL }\fi

\bibitem[{Adams et~al.(1988)Adams, Balas, and Zawack}]{Adams1988LA01-40}
Adams, J., Balas, E., Zawack, D., 1988. The shifting bottleneck procedure for
  job shop scheduling. Management Science 34~(3), 391--401.

\bibitem[{Aiex et~al.(2003)Aiex, Binato, and Resende}]{Aiex2003GRASPwithPR}
Aiex, R.~M., Binato, S., Resende, M. G.~C., 2003. Parallel {GRASP} with
  path-relinking for job shop scheduling. Parallel Computing 29~(4), 393 --430.

\bibitem[{Applegate and Cook(1991)}]{ApplegateSpring1991ORB01-10}
Applegate, D., Cook, W., 1991. A computational study of the job-shop scheduling
  problem. ORSA Journal on Computing 3~(2), 149--156.

\bibitem[{Balas and Vazacopoulos(1998)}]{Balas1998GLS}
Balas, E., Vazacopoulos, A., 1998. Guided local search with shifting bottleneck
  for job shop scheduling. Management Science 44~(2), 262--275.

\bibitem[{Beck et~al.(2011)Beck, Feng, and Watson}]{Beck2011CP/LS}
Beck, J.~C., Feng, T.~K., Watson, J.-P., 2011. Combining constraint programming
  and local search for job-shop scheduling. Informs Journal on Computing
  23~(1), 1--14.

\bibitem[{Cheng et~al.(2013)Cheng, Peng, and L{\"u}}]{Cheng2013HEA}
Cheng, T. C.~E., Peng, B., L{\"u}, Z., 2013. A hybrid evolutionary algorithm to
  solve the job shop scheduling problem. Annals of Operations Research, 1--15.

\bibitem[{Demirkol et~al.(1997)Demirkol, Mehta, and
  Uzsoy}]{Demirkol1997DMU01-80}
Demirkol, E., Mehta, S., Uzsoy, R., 1997. A computational study of shifting
  bottleneck procedures for shop scheduling problems. Journal of Heuristics 3,
  111--137.

\bibitem[{Fisher and Thompson(1963)}]{Fisher1963FT06-20}
Fisher, H., Thompson, G.~L., 1963. Probabilistic learning combinations of local
  job-shop scheduling rules. Industrial scheduling, 225--251.

\bibitem[{Gon{\c{c}}alves and Resende(2013)}]{Goncalves2013BRKGA}
Gon{\c{c}}alves, J.~F., Resende, M. G.~C., 2013. An extended akers graphical
  method with a biased random-key genetic algorithm for job-shop scheduling.
  International Transactions in Operational Research.

\bibitem[{Lawrence(1984)}]{Lawrence1984LA01-40}
Lawrence, S., 1984. Resource constrained project scheduling: an experimental
  investigation of heuristic scheduling techniques (supplement). Tech. rep.,
  Graduate School of Industrial Administration, Carnegie Mellon University,
  Pittsburg.

\bibitem[{Nagata and Tojo(2009)}]{Nagata2009GES}
Nagata, Y., Tojo, S., 2009. Guided ejection search for the job shop scheduling
  problem. Evolutionary Computation in Combinatorial Optimization, 168--179.

\bibitem[{Nasiri and Kianfar(2012{\natexlab{a}})}]{Nasiri2012GES/TS}
Nasiri, M.~M., Kianfar, F., 2012{\natexlab{a}}. A {GES}/{TS} algorithm for the
  job shop scheduling. Computers \& Industrial Engineering 62~(4), 946--952.

\bibitem[{Nasiri and Kianfar(2012{\natexlab{b}})}]{Nasiri2012guidedTSPR}
Nasiri, M.~M., Kianfar, F., 2012{\natexlab{b}}. A guided tabu search/path
  relinking algorithm for the job shop problem. The International Journal of
  Advanced Manufacturing Technology 58~(9-12), 1105--1113.

\bibitem[{Nowicki and Smutnicki(1996)}]{Nowicki1996TSAB}
Nowicki, E., Smutnicki, C., 1996. A fast taboo search algorithm for the job
  shop problem. Management Science 42~(6), 797--813.

\bibitem[{Nowicki and Smutnicki(2005)}]{Nowicki2005i-TSAB}
Nowicki, E., Smutnicki, C., 2005. An advanced tabu search algorithm for the job
  shop problem. Journal of Scheduling 8, 145--159.

\bibitem[{Pardalos and Shylo(2006)}]{Pardalos2006GES}
Pardalos, P.~M., Shylo, O.~V., 2006. An algorithm for the job shop scheduling
  problem based on global equilibrium search techniques. Computational
  Management Science 3, 331--348.

\bibitem[{Pardalos et~al.(2010)Pardalos, Shylo, and
  Vazacopoulos}]{Pardalos2010AlgFix}
Pardalos, P.~M., Shylo, O.~V., Vazacopoulos, A., 2010. Solving job shop
  scheduling problems utilizing the properties of backbone and ¡°big valley¡±.
  Computational Optimization and Applications 47, 61--76.

\bibitem[{Qing-dao-er ji and Wang(2012)}]{Qing-dao-er-ji2012HGA}
Qing-dao-er ji, R., Wang, Y., 2012. A new hybrid genetic algorithm for job shop
  scheduling problem. Computers \& Operations Research 39~(10), 2291--2299.

\bibitem[{Shen and Buscher(2012)}]{Shen201214}
Shen, L., Buscher, U., 2012. Solving the serial batching problem in job shop
  manufacturing systems. European Journal of Operational Research 221~(1), 14
  -- 26.

\bibitem[{Storer et~al.(1992)Storer, Wu, and Vaccari}]{Storer1992SWV01-15}
Storer, R.~H., Wu, S.~D., Vaccari, R., 1992. New search spaces for sequencing
  problems with application to job shop scheduling. Management Science 38~(10),
  1495--1509.

\bibitem[{Taillard(1994)}]{TaillardSpring1994TA01-50}
Taillard, E.~D., 1994. Parallel taboo search techniques for the job shop
  scheduling problem. ORSA Journal on Computing 6~(2), 108--117.

\bibitem[{Yamada and Nakano(1992)}]{Yamada1992YN1-4}
Yamada, T., Nakano, R., 1992. A genetic algorithm applicable to large-scale
  job-shop problems. In: PPSN. pp. 283--292.

\bibitem[{Zhang et~al.(2007)Zhang, Li, Guan, and Rao}]{Zhang2007TS}
Zhang, C.~Y., Li, P.~G., Guan, Z.~L., Rao, Y.~Q., 2007. A tabu search algorithm
  with a new neighborhood structure for the job shop scheduling problem.
  Computers \& Operations Research 34~(11), 3229 -- 3242.

\bibitem[{Zhang et~al.(2008)Zhang, Li, Rao, and Guan}]{Zhang2008TSSA}
Zhang, C.~Y., Li, P.~G., Rao, Y.~Q., Guan, Z.~L., 2008. A very fast {TS}/{SA}
  algorithm for the job shop scheduling problem. Computers \& Operations
  Research 35~(1), 282-- 294.

\end{thebibliography}

\end{document}